\documentclass{article}
\usepackage{indentfirst}
\usepackage{float}
\usepackage{changes}
\usepackage{array}

\begin{document}

\title{An Overview of Recent Solutions to and Lower Bounds for the Firing Synchronization Problem}
\author{Thiago Correa, Breno Gustavo, Lucas Lemos, Amber Settle}
\date{\today}
\maketitle

\section{Introduction}
Complex systems in a wide variety of areas such as biological modeling, image processing, and language recognition can be modeled using networks of very simple machines called finite automata. A finite automaton is an idealized machine that follows a fixed set of rules and changes its state to recognize a pattern, accepting or rejecting the input based on the state in which it finds itself at the conclusion of the computation. In general a subsystem can be modeled using a finite automaton if it can be described by a finite set of states and rules. Connecting subsystems modeled using finite automata into a network allows for more computational power. One such network, called a cellular automaton, consists of an $n$-dimensional array for $n > 1$  with a single finite automaton located at each point of the array. Each automaton reads the current state of its neighbors to determine how to modify its own state. The interest in studying cellular automata is not in the description of the local state transitions of each machine, but rather in studying the global behavior across the entire array.

One of the oldest problems associated with cellular automata is the firing synchronization problem, originally proposed by John Myhill in 1957. In the firing synchronization problem each machine of the cellular automata is identical, except for a single designated machine called the general or initiator.  The synchronization starts with every cell in a \textit{quiescent} state, except the initiator which starts in the \textit{general} state and is the one initiating the synchronization. The next state of each cell is determined by the cell's state and the state of its neighbors, with a required rule being that a machine that is quiescent and has two neighbors that are quiescient must remain so in the next round of the computation. The goal of the problem is to synchronize the array of automata, which means that each machine enters a unique designated \textit{firing} state for the first time and simultaneously at some point during the execution.
    
    
As might be expected with a problem that has existed as long as the firing synchronization problem, there are a wide range of variants for the network and its initial configuration. In the original problem the initiator is placed at one end of the array, either on the left or the right. A problem that is more complex than the original problem is the generalized problem, defined by Moore and Langdon in 1968 \cite{MooreLangdon1968}. In this version the initiator can be placed anywhere in the array, although there is a restriction that there is a single initiator. The problem can be made more complex by requiring that the solution be symmetric, a requirement first introduced by Szwerinski in 1982 \cite{szwerinski1982time}. In this type of solution an automaton cannot distinguish its right and left neighbors, eliminating any directional information provided to the automaton and making the problem more difficult to solve.

Another set of variants of the problem consider alterations to the underlying structure of the network. All of the problems described above operate on a one-dimensional array, which is the simplest configuration for a cellular automaton. In that configuration the end machines do not have neighbors, so that the end machines make state transitions based only on their own state and one existing neighbor. An alternate approach in one dimension is to connect the two end machines to each other, forming a ring rather than an array. The network can also be expanded to more than one dimension, producing a grid (two dimensions), a cube (three dimensions), or more generally a hybercube (four or more dimensions).

Once a variant and configuration for the network has been established, it is important to measure the complexity of the solution to the problem. One measure of a solution is the time in which the solution causes the network to synchronize. A minimal-time solution is one that synchronizes the network as quickly as possible, and a non-minimal-time solution is one that synchronizes the network at some point but requires more steps/rounds than the minimal-time solution. For example, the minimal time required for the original problem on a one-dimensional array is $2n - 2$ steps for an array with $n$ machines. Any solution sychronizing in $2n - 2$ steps is minimal-time, while those requiring more than $2n - 2$ rounds are non-minimal-time solutions. Solutions can also be compared based on the number of states for each finite automaton, with solutions that require fewer states seen as more difficult and therefore better. Another way to compare solutions was introduced by Vollmar in 1982 \cite{vollmar1982some}. Vollmar's metric is the state complexity of a solution, which is the combination of the number of states and number of transitions required by the solution.

As with any long-standing problem, there are a large number of solutions to the firing synchronization problem. Our goal, and the contribution of this work, is to summarize recent solutions to the problem. We focus primarily on solutions to the original problem, that is, the problem where the network is a one-dimensional array and there is a single initiator located at one of the ends. We summarize both minimal-time and non-minimal-time solutions, with an emphasis on solutions that were published after 1998. We also focus on solutions that minimize the number of states required by the finite automata. In the process we also identify open problems that remain in terms of finding minimal-state solutions to the firing synchronization problem.
    
\section{Early results}



While the focus of this paper is on recent solutions to and lower bounds for the firing synchronization problem (FSP), it is useful to summarize earlier results to provide context for more recent ones. In this section we summarize results providing solutions to the FSP using a (relatively) small number of states as well as known lower bounds on the number of states required to solve the FSP. This section focuses on solutions and lower bounds published prior to 1999.

\subsection{Minimal-time results}

The first minimal-time solution to the original one-dimensional FSP was produced in 1962 when Goto gave a solution with over a thousand states \cite{Goto1962}. Once it became clear that a minimal-time solution was possible work quickly turned to minimizing the number of states required by the solution. In 1966 Waksman \cite{waksman1966optimum} gave a 16-state minimal-time solution, and Balzer \cite{Balzer1967} independently produced an 8-state solution using the same ideas. In 1987 Mazoyer produced a 6-state solution to a restricted version of the problem in which the initiator is always located at the left endpoint of the array \cite{Mazoyer1987}.

In 1968 Moore and Langdon introduced the generalized problem in which the initiator can be located anywhere in the array and gave a 17-state minimal-time solution \cite{MooreLangdon1968}. An improvement on this solution was produced in 1970 with the publication of a 10-state solution \cite{VMP1970}. Further work was done by Szwerinski who produced a 10-state symmetric solution \cite{szwerinski1982time}. Szwerinski's solution was subsequently improved to a 9-state solution \cite{settle1999new}.

There were many fewer early solutions for the firing synchronization problem on a ring. In 1989 Culik modified Waksman's solution to function on the one-dimensional ring \cite{Culik1989}. Like Waksman's solution Culik's ring solution uses 16 states.

\subsection{Non-minimal-time solutions}
An early non-minimal-time solution to the FSP was created by Fischer who gave a 15-state solution that synchronized an array of $n$ automata in 3$n$-4 steps\cite{Fischer:1965:GPO:321281.321290}. Later
Mazoyer suggested that all solutions with few states must necessarily be minimal-time, a conjecture based on the idea that the simplest solution will naturally be the fastest. Yun\'{e}s disproved the conjecture in 1994 by giving an implementation of a divide and conquer solution originally given by McCarthy and Minsky \cite{MM1972} that requires only 7 states and synchronizes in time $t(n) = 3n \pm 2 \theta_{n}logn+C$ where $0 \leq \theta_{n} < 1$ \cite{Yunes1994}. The solution is for the restricted version of the original problem where the initiator must be located at the left endpoint. A 6-state solution was produced in 1999 that allows the initiator to be location at either the left or right endpoint \cite{settle1999new}. The 6-state solution synchronizes $n$ automata in $2n - 1$ rounds if the initiator is located at the left endpoint and $3n+1$ rounds if the initiator is located at the right endpoint.

The same authors produced a 7-state non-minimal-time solution to the generalized problem \cite {settle1999new}. The 7-state solution requires $2n-2+k$ rounds to synchronize an array of length $n$ when the initiator is located in position $k$ of the array.

Also in that work is an 11-state non-minimal-time solution to the ring variant of the FSP \cite{settle1999new}. To synchronize a ring of length $n$ the solution requires $2n-3$ time steps if n is even and $2n-4$ time steps if n is odd.

\subsection{Lower bounds}
Early work on finding lower bounds for the number of states for solutions to the firing synchronization problem is limited. The earliest state lower bound for the FSP was claimed in 1967 by Balzer \cite{Balzer1967} and confirmed by Sanders in 1994 \cite{Sanders1994}. Sanders showed that there is no 4-state minimal-time solution to the restricted original problem where the initiator must be located at the right endpoint of the array. Sander's (and Balzer's) technique involved a modified exhaustive search in which a program examined all possible 4-state solutions to demonstrate that none of them correctly solved the problem. It is crucial for the programs that only minimal-time solutions are examined since it allows the program to discard any solution that has not achieved synchronization of $n$ automata by time step $2n-2$.

The first analytical proofs of state lower bounds came in the late 1990s. One result showed that there is no minimal-time 3-state solution to the original FSP \cite{settle1999new}. A similar approach was able to show that there are also no 3-state non-minimal time solutions to the problem \cite{settle1999new}. Weaker results are produced for 4-state and 5-state solutions, showing that there are no solutions subject to a small number of constraints on the solution \cite{settle1999new}.

A less elegant exhaustive case analysis showed that there is no 4-state minimal-time solution to the ring variant of the firing synchronization problem under certain restricted conditions \cite{settle1999new}.

\section{Summary of recent research}
The purpose of this paper is to document and classify recent results about the firing synchronization problem. Our main focus is on results published between 1998 and 2015. In this section we first describe the methodology we used to find the recent results. We then describe each of the results, categorizing them by type of problem and type of solution.

\subsection{Methodology}
In order to summarize current research on minimal-state solutions to the firing synchronization problem we conducted a literature review. We began by identifying pre-1998 authors who had published on the problem and from that the conferences and journals in which their work had been published. That resulted in the following list of conferences and journals:
\begin{itemize}
\item Journals
	\begin{itemize}
		\item Information and Computation
		\item Information and Control
		\item Information Processing Letters
		\item International Journal of Unconventional Computing
		\item Journal of Algorithms
		\item Journal of Computer and System Sciences
		\item Parallel Processing Letters
		\item SIAM Journal on Computing
		\item Theoretical Computer Science
	\end{itemize}
\item Conferences
	\begin{itemize}
		\item International Colloquium on Structural Information and Communication Complexity (SIROCCO)
		\item International Conference on Cellular Automata for Research and Industry (ACRI)
		\item International Conference on Membrane Computing
		\item International Conference on Networking and Computing (ICNC)
		\item international Workshop on Cellular Automata and Discrete Complex Systems
		\item Symposium of Principles of Distributed Computing
	\end{itemize}
\end{itemize}
We then examined all of the journals and conferences looking for articles with relevant titles between 1998 and 2015. Among the journals only Information Processing Letters, Parallel Processing Letters, and Theoretical Computer Science yielded articles in the specified time period. In the specified time period only the International Colloquium on Structural Information and Communication Complexity (SIROCCO) and the International Conference on Cellular Automata for Research and Industry (ACRI) yielded relevant articles.

We also searched for more recent articles written by pre-1998 authors. When we were aware of them, we included papers that had referenced pre-1998 articles. Any article found via any means also added to the list of authors included in the search. We also did searches for papers including relevant keywords, such as cellular automata, firing synchronization, and firing squad.

In the remainder of the paper we summarize the articles containing results that minimize the number of states used to solve the firing synchronization problem. While our literature review may have missed some articles, either because they were published in venues of which we were unaware or by authors who were not in any way connected to previous results, we believe that the majority of articles containing state minimization results are represented.

\subsection{Background information}
In order to understand some of the results discussed later, some background information is necessary. We describe in this section Wolfram's rules as well as discuss a solution technique common to many non-minimal-time results.

\subsubsection{Wolfram's rules}\label{sec:wolfram}
Stephen Wolfram is a scientist known for his work on cellular automaton and his contribution to theoretical physics. In 2002 he published \emph{A New Kind of Science}\cite{Wolfram2002} in which he presents systems he calls \emph{simple programs}. Generally these simple programs have a very simple abstract framework like simple cellular automata, Turing machines, and combinators. The focus of \emph{A New Kind of Science} is to use these simple programs to study simple abstract rules, which are, essentially, elementary computer programs.

In our survey, we found solutions based on some of Wolfram's rules. Here we describe the basic structure of the rules and discuss two of them in particular. In Wolfram's text elementary cellular automata have two possible values for each cell which are colored white (for 0) or black (for 1). The rules for a transition are based on triples, with the next color of a cell depending on its color and the colors of its two immediate neighbors. The configuration of the transition rules are always the same and are given in the following order:
\begin{enumerate}
\item three black squares (111)
\item two black squares followed by a white square (110)
\item a black square, a white square, and a black square (101)
\item a black square, and two white squares (100)
\item a white square and two black squares (011)
\item a white square, a black square, and a white square (010)
\item two white squares and a black square (001)
\item three white squares (000)
\end{enumerate}

Thus each set of transition rules can be thought of as an eight-digit binary number since each rule needs to transition either to white (0) or black (1). 
The rules are named based on the representation of the state produced in the automaton in binary. So rule 60 is the one that assigns the transitions the binary value 00111100. This means that it defines the transitions as 111 $\rightarrow$ 0, 110 $\rightarrow$ 0, 101 $\rightarrow$ 1, 100 $\rightarrow$ 1, 011 $\rightarrow$ 1, 010 $\rightarrow$ 1, 001 $\rightarrow$ 0, and 000 $\rightarrow$ 0.  

Another Wolfram's rule mentioned in results we discuss later in this paper is rule 150 which specifies that the states above transition to the state encoded by the binary representation of the number 150 (binary value 10010110)

\subsubsection{3n-step algorithm} \label{ssec:3nstep}

The 3n-step algorithm is an interesting approach to synchronizing a two-dimensional cellular automaton due to its simplicity and straightforwardness. This type of approach is used when a minimal-time solution is not required.

In the design of any 3n-step approach the most crucial step is to find the center cell of the array to be synchronized. The basic mechanism for doing this is to use two signals moving at different speeds. The first signal is called \textit{a-signal} and it moves at the speed of one cell per unit step. The second signal is called \textit{b-signal} and it moves at a speed of one cell per three unit steps. When the \textit{a-signal} finds the end of the array it returns to meet the \textit{b-signal} at the half of the array, but the reflected \textit{a-signal} is called \textit{r-signal}. If the length of the array is odd, the cell C$_{\lceil n/2 \rceil}$ becomes the general, if the length is even two generals are created and each general is responsible for synchronizing it is left and right halves of the cellular space. 



Recursion is an important part of this algorithm. After finding the center cell, the process has to be started over with the new general, or generals if it is an even array. And the simplest way to start over is to apply recursion on the left and right side of the now divided array. After many steps of recursion, which depends on the size of the array, the problem is reduced to a problem of size two which is the last step before firing.

\subsection{Overview of results}
In the remainder of the paper we describe and categorize results for the FSP, with a focus on results published between 1998 and 2015. To provide an overview of the work we summarize the table below gives information about the results discussed.

\begin{table}[H]
\centering
\caption{Summary of results for the FSP}
\label{ovvtable} 
\begin{tabular}{|>{\centering\arraybackslash}p{2.0cm}|c|>{\centering\arraybackslash}p{2.0cm}|c|c|>{\centering\arraybackslash}p{1.8cm}|c|c|}\hline

\multicolumn{1}{|>{\centering\arraybackslash}p{1.8cm}|}{Author} & 
\multicolumn{1}{>{\centering\arraybackslash}p{1.3cm}|}{Year} & 
\multicolumn{1}{>{\centering\arraybackslash}p{2.0cm}|}{Algorithm} & 
\multicolumn{1}{>{\centering\arraybackslash}p{1.4cm}|}{\# States} & 
\multicolumn{1}{>{\centering\arraybackslash}p{1.3cm}|}{\# Rules} & 
\multicolumn{1}{>{\centering\arraybackslash}p{1.8cm}|}{Time} & 
\multicolumn{1}{>{\centering\arraybackslash}p{2cm}|}{General's position} & 
\multicolumn{1}{>{\centering\arraybackslash}p{2.5cm}|}{Notes}\\ \hline

Settle \cite{settle1999new} & 1999 & Szwerinski & 9 & - & $2n-1$ & arbitrary & symmetric\\ \hline

Settle \cite{settle1999new} & 1999 & Mazoyer & 6 & 134 & $3n+1$ & left or right & -\\ \hline

Settle \cite{settle1999new} & 1999 & Mazoyer + $3n$-step & 7 & 127 & $2n-2+k$ & arbitrary & -\\ \hline

Noguchi \cite{Noguchi04} & 2004 & Balzer & 8 & 119 & $2n-2$ & left & - \\ \hline

Umeo, Maeda and Hongyo \cite{umeo2006design} & 2006 & $3n$-step & 6 & 115 & max(k, n − k + 1) + 2n+ O(log n) & arbitrary & symmetric\\ \hline

Umeo and Yanagihara \cite{umeo2009small} & 2009 & $3n$-step & 5 & 71 & $3n-3$ & left or right & $n=k^2$\\ \hline

Umeo, Yunes and Kamikawa \cite{umeo20084} & 2008 & Wolfram's rule 60 & 4 & - & $2n - 2$ & left & $N=2^n$\\ \hline

Umeo, Yun\`{e}s and Kamikawa \cite{umeo20084} & 2008 & Wolfram's rule 150 & 4 & - & $2n-2$ & left & $N=2^n+1$\\ \hline

Yun\`{e}s \cite{yunes20084} & 2008 & Wolfram's rule 60 & 4 & 32* & $2^{n+1}-1$ & left & $N=2^n$\\ \hline

Umeo, Yun\`{e}s and Kamikawa \cite{umeo2009family} & 2009 & Computer-based investigation  & 4 & - & $2n - 2$ & left & $n=k^2$, symmetric\\ \hline

Umeo, Kamikawa and Nishioka \cite{umeo2010generalized} & 2010 & Computer-based investigation & 8 & 222 & $2n - 2$ & arbitrary & asymmetric \\ \hline


\end{tabular}
\end{table}

In the table we list the author for the publication from which the results are drawn, the year in which the publication appears, the type of algorithm employed (described either by citing the author who first conceived it or the general approach taken), the number of states used in the solution, the number of rules (e.g. number of transitions) required by the solution, the time required to synchronize an array of $n$ automata, where the initiator is located when the synchronization is begun, and any important notes about the solution.

\subsection{Minimal-time solutions}
Recall that a minimal-time solution is one that takes $2n-2$ steps to synchronize an array of $n$ automata. Each section below summarizes a paper by a set of authors providing a minimal-time solution to the FSP.

\subsubsection{Settle and Simon: 2002}
Settle and Simon \cite{settle2002smaller} provide a range of different solutions to the firing synchronization problem. The first solution is a minimal-time solution to the generalized problem, where the initiator can be located in any cell in the array. The 9-state minimal-time to the generalized problem is an improvement from the 10-states solution created by Szwerinski\cite{szwerinski1982time}. One common strategy to deal with problems that use the generalized problem is first make the generalized problem look like the original problem, and then solve the original problem. Szwerinski's solution needed two states to transform the generalized problem into the original problem, and Settle and Simon and improved that to use only one more state.

The second contribution of the paper is a 6-states non-minimal-time solution to the original problem, where the initiator must be on either the left or right side of the array.
The 6-state automaton is based on Mazoyer's 6-state solution, but the version by Settle and Simon the initiator may be on either the left and right side of the array. Mazoyer's solution works by dividing the array in unequal parts of $2/3n$ and $1/3n$ to create subproblems that will also be divided recursively, and Settle and Simon's solution works the same way.

The third solution in the paper is a 7-state non-minimal-time solution to the generalized problem. The solution also based on the optimum time 6-state solution from Mazoyer. One state is added to the Mazoyer solution to transform it into a solution to the generalized problem. To do this Settle and Simon used the 3n-Step algorithm to allow the initiator to be put in any place of the array. The new state is used to find the middle of the array, from which point the work continues with the Mazoyer solution.

\subsubsection{Noguchi: 2004}
The focus of Noguchi's work is to provide a more straightforward solution with fewer rules to the 8-state minimal time problem on rings\cite{Noguchi04}. His 8-state solution uses a strategy inspired by Waksman\cite{waksman1966optimum} and Balzer\cite{Balzer1967} where waves are used to gather, store, and pass information about the system, typically by halving the array and placing initiators at the midpoints. This is not a solution to the generalized problem, so the general must be at the beginning of the line. The strategy used in the solution has a main wave that travels at full speed, and a reverse wave that goes back also at maximum speed. These waves are used to detect the middle point, quarter point, and so on of the line. Doing that reduces the problem to sub-problems of the original problem, and every sub-problem can also be reduced recursively. For every sub-problem, a new general and consequently a new main wave and reverse wave are created.

The algorithm works sending multiple waves in different speeds. The primary wave travels a one cell per time, and after the primary wave, multiple middle waves are sent. This multiple waves will be responsible for the creation of the future middle points. The n-middle waves that are sent after the primary wave moves one machine in the time that the primary wave moves $2^n$ machines. This will ensure that when the primary wave reflects it will find the middle waves in the respective middle spots of the array, $1/2$, $1/4$, $1/16$, and so on. The primary wave also is responsible for sending backward waves. These backward waves are responsible for passing-through the middle waves. This contact serves as instruction for the middle waves to start the process again. 

\subsubsection{Umeo, Yun\`{e}s, and Kamikawa: 2008}
In the work by Umeo, Yun\`{e}s and Kamikawa\cite{umeo20084}, some elements of a new family of time-optimal solutions to a less restrictive firing squad synchronization problem are presented. The solutions are based on elementary algebraic cellular automata. The authors present some 4-state solutions which synchronize every line whose length is $2^n$ or $2^n + 1$ based on Wolfram's rules 60 and 150 described in Section~\ref{sec:wolfram}. Umeo, Yun\`{e}s and Kamikawa were able to construct different 4-state solutions to the FSP which synchronizes every line of length $2^n$ or $2^n + 1$. Three of these solutions use only 33 transitions and one uses only 30 transitions.

Rule 60 when run on a configuration of length a power of two where the left end cell is 1 (black) led to something that looked like a synchronization, although it should be noted that rule 60 is not a solution to the problem. So to create something able to synchronize every power of two the authors use a simple folding of the space-time diagram 
of rule 60 running on a line of length $2^{n+1}$, resulting in a space-time diagram of a 4-state automata running on a line of length $2^n$. Using this modification they could obtain the synchronization at the time $2n$ for a line of length $n$. Using the same concept, they also present a solution that synchronizes every line of length $N = 2^n + 1$ at time $2N - 1$ ($2N - 2$ steps), so that it is a strict optimum-time solution.

They also noted that the number of non quiescent cells is not in the order of $n log(n)$ for a line of length $n$, neither it is in the order of $n^2$, but something in between.

\subsubsection{Umeo, Kamikawa and Yun\`{e}s: 2009}
In this paper Umeo, Kamikawa and Yun\`{e}s\cite{umeo2009family} provide a partial solution to the FSP by presenting a family of 4-state solutions to synchronize any one-dimensional ring of length $n = 2^k$ where $k$ represents any positive integer.

The authors consider only symmetrical 4-state solutions for the ring. In their approach they use a computer to search the transition rule set in order to find a FSP solution. They did this by generating a symmetrical 4-state transition table and computing the configuration of each transition table for a ring of length $n$. They assume that $Q$ is a quiescent state, $A$ and $G$ are auxiliary states and $F$ is the firing state. Their program starts from an initial configuration: $G\overbrace{Q,...Q}^{n-1}$ where $2 \le n \le 32$ and checks if each transition table yields a synchronized configuration: $\overbrace{FF,...F}^n$ during the time $t$ where $n \le t \le 4n$ and the state $F$ never appears before that. By doing this they found that there were 6412 successful synchronized configurations. Most of them included redundant entries, so they removed the redundancies and compared and classified the valid solutions into small groups. After that process they obtained seventeen solutions: four optimum-time solutions, four nearly-optimum time solutions, and nine non-optimum time solutions. All of these seventeen solutions can synchronize rings of length $n = 2^k$ for any positive integer $k$.

They also converted the solutions into solutions for an open ring, that is, a conventional one-dimensional array. They found that the converted 4-state protocol can synchronize any one-dimensional array of length $n = 2^k + 1$ with the left-end general both in state $G$ and $A$ in optimum $2n -2$ steps, for any positive integer $k \ge 1$.

\subsubsection{Umeo, Kamikawa, Nishioka and Akiguchi: 2010}

\indent In this paper the authors Umeo, Kamikawa, Nishioka and Akiguchi \cite{umeo2010generalized} present a computer study on different solutions to the generalized firing synchronization problem (GFSP). Recall that the GFSP may be described as the original problem with the general on the far left or right of the array at time $t = -(k - 1)$, where $k$ is the number of cells between the general and the nearest end. This study reveals inaccuracies in previous solutions, such as redundant rules and unsuccessful synchronizations. The paper also introduces a new eight-state solution to the GFSP, which surpasses the previous best solution that had nine states. Their use of a computer-assisted approach helped the transition table of their new solution to not have flaws in redundancy or unsuccessful synchronizations. A six-state solution is also examined in the study.

The study presented in this paper takes into account the state transition tables for each solution of the GFSP being analyzed. The first transition table studied is Moore and Langdon's 17-state optimum-time solution \cite{MooreLangdon1968}. This solution had problems synchronizing a relatively large number of arrays with several positions of the initial general. The second transition table studied is Varshavsky, Marakhovsky and Peschansky's 10-state optimum time solution \cite{VMP1970}. This solution also has unsuccessful synchronizations. The third table studied is Szwerinski's optimum-time ten-state solution \cite{szwerinski1982time}. This solution had no errors in the table, so it didn't present any unsuccessful synchronization; however, it had 21 redundant rules.  The fourth table studied is Settle and Simon's 9-state optimum time solution \cite{settle2002smaller}. This table had no errors, but it presented 16 redundant rules. The 8-state optimum-time solution is then presented. This solution has 222 rules none of them being redundant. The last table studied is Umeo, Maeda and Hongyo's 6-state non-optimum-time solution \cite{umeo2006design}. This solution has 115 states none of them being redundant and is considered a 3n-step solution.

\subsection{Non-minimal-time solutions}
Recall that a non-minimal-time solution is one that takes more than $2n-2$ steps to synchronize an array of $n$ automata. Each section below summarizes a paper by a set of authors providing a non-minimal-time solution to the FSP.

\subsubsection{Umeo, Maeda and Hongyo: 2006} 
Umeo, Maeda and Hongyo \cite{umeo2006design} give a new 3n-step algorithm that improves the lower bound of the generalized firing synchronization to a 6-state symmetric solution. The paper has two parts. In the first part it provides a 6-state solution to the original problem with the initiator on the left side. In the second part it gives a 6-state solution to the generalized problem, where the initiator can be placed anywhere on the array. The main difference between the two solutions is that on the second one, more rules are used to transform the solution to the original problem on the solution to the generalized problem.

This solution like other that uses the 3n - step algorithm starts with the propagation of the a and b signals, on this solution the propagation of the a- signal is represented by the state P. While P is going away on the right direction at speed of one cell per time, it takes a a state R and M alternatively at each step until either the b-signal or the r-signal arrives at the cell itself. The b-signal is represented by the propagation of a 1/3 speed signal where the cells take a state R, R and Z for each three steps. And finally, the R signal is represented as a 1/1 speed signal of the Z state.   

One of the key ideas used on this paper to improve the 3n-step algorithm is based on the use of the quiescent cells of the zone T. The zone T is the triangle area circled by a-, b-, and r-signals in the time-space diagram. In the implementations of Fischer and Yun\`{e}s, all cells in zone T keep quiescent state and are always inactive during the computations. These authors use a strategy that depends on making all cells inside the zone T active. Using the quiescent cells from this zone the authors successfully reduced the number of states by reusing the Q states to help the r- signal (Z state) find the center of the array.

Finally, note that the result presented in this paper is an improvement of Yun\`{e}'s 7-state solution. Further, the 6-state solution is the smallest one known at present in the class of non-trivial 3n-step synchronization algorithms. The authors also achieved a increase in the number of working cells from $O(n log n)$ to $O(n2)$, and with that a state-efficient synchronization algorithm.

\subsubsection{Umeo and Yanagihara: 2007}

Here Umeo and Yanagihara \cite{umeo2007smallest} provide a partial 5-state solution to the firing synchronization problem. Using a 3n-step algorithm, this solution can synchronize any one-dimensional cellular array of length $n=2^k$ in 3n - 3 steps, for any positive integer k and with the initiator positioned on the right side. The first solution using this algorithm was from Fischer\cite{Fischer:1965:GPO:321281.321290} who gave a 15-state implementation.

This 5-state solution is a small but partial solution to the original problem, since it has some limitations regarding the length of the array n. Other authors like Settle and Simon\cite{settle2002smaller} and Umeo\cite{umeo2006state} provided complete solutions to the generalized problem using the same kind of algorithm but with a higher number of states. Settle and Simon provided a solution using only 7 states\cite{settle2002smaller}, and Umeo did using 6 states\cite{umeo2006state}. Unlike this one in both solutions the initiator can be placed in any part of the array. 

The first two states of the solution are used to create the ripple drivers that enable the propagation of the b- signal at 1/3 speed (state S). The a- signal is also realized using the first two states (state R). Every two steps the a- signal generates a 1/1 speed signal in state Q, this signal is transmitted in the reverse direction.
Using the reverse State Q, a third state S is added that will be responsible for the b- signal. Each ripple driver can be used to drive the propagation of the b-signal to its right neighbor. The three-step separation of two consecutive ripples enables the b-signal to propagate at 1/3 speed. Finally, state L is used for a reflected R-signal. The return signal propagates left at 1/1 speed. Any cell where the return signal passes remains in a quiescent state. At time t=3n/2, the b-signal and the return signal meet at $Cn/2$ and $Cn/2+1$. At the next step, the cells $Cn/2$ and $Cn/2+1$ take a state L and R, and these states act like generals for the left and right half of the array, and the process starts over again recursively.

\subsubsection{Yun\`{e}s: 2008} \label{ssec:num1}
In this paper Yun\`{e}s\cite{yunes20084} presents a solution to the FSP based on Wolfram's rule 60 discussed in Section~\ref{sec:wolfram}. He accomplishes his result using an algebraic approach instead of geometrical constructions. This solution solves the problem on an infinite number of lines but not all possible lines. Its state complexity is the lowest possible (4 states and 32 transitions).

As mentioned before, running the rule 60 on a configuration where the left end cell is 1 (black) leads to something that looks like synchronization of lines of length which are powers of two, but it is not a solution to the problem. Yun\`{e}s points that whatever the solution, if it synchronizes an infinite number of lines then for some $N$ the synchronization can only appear at time $2n -n$ for a line of length $n > N$. Then he designed the algorithm so that rule 60 ran sending a signal wave 1 from the left most cell. When the leading 1 reaches the right end cell, another symmetric rule 60 is launched. By doing that, a property of the global function is exploited so that the full interleaving of two basic configurations is reached and the synchronization appears naturally.

This leads to the paper's main theorem: There exists a 4-state solution to the firing squad synchronization problem which synchronizes all lines of length a power of 2 in $2^{n+1}-1$ steps. Furthermore the number of non-quiescent cells in the space-time diagram of a line of length $m = 2^n$ is $3 \cdot m^{\log{(3)}}$. They show that this theorem is optimal by proving that it is impossible to synchronize any line of length $n \ge 5$ with only 3 states. Using only 32 program instructions makes this solution very simple. Previous results most often used geometrical constructions, and this was the first time an algebraic approach produced a solution for the FSP.

\subsubsection{Umeo and Yanagihara: 2009}

Umeo and Yanagihara present in this paper a partial solution to the firing synchronization problem with 5 states\cite{umeo2009small}. This solution only works with arrays of length $n = 2^k$, and it takes $3n - 3$ steps to synchronize, where $k$ is a positive integer. This solution uses a $3n$-step algorithm, which was explained in detail in Section \ref{ssec:3nstep}. The advantage of the use of such algorithm is its simplicity, which makes the approach easily understandable. 

The internal set of 5 states for this solution is represented by $\mathcal{Q} = \{Q,L,R,S,F\}$. First the authors used a 2-state implementation for the wave, which implements the a-signal and enables the future propagation of the b-signal. The 2-states are $\{Q,R\}$. The states $\{Q,R,S\}$ implement the a- and b-signals which were explained in the subsection \ref{ssec:3nstep}. The state $L$ was used to implement the search for the center cells. After all the center cells, for sections and subsections of the cellular space, were found they could achieve the 
firing state $F$.

The solution described in this paper is proposed as not the smallest solution to the problem but interesting in its own way. The authors achieved a 5-state partial solution, since this solution only works for a certain set of arrays. Because of that we can consider it a small partial solution for the FSP. However, it is not the smallest, since a 4-state partial solution, presented in Section~\ref{ssec:num1}  also exists.

\subsection{Surveys and generalized problems}
In this section we summarize a survey article on the FSP and discuss an article that considers a solution for the FSP on a two-dimensional array. The two dimensional array is a cellular automaton composed of an array of m × n cells. The state of any cell is not only influenced by the state of the cell on the both sides, but also the cells at north and south. Several synchronization algorithms on two-dimensional arrays have been proposed, including Grasselli \cite{grasselli1975synchronization}, Kobayashi \cite{kobayashi1977firing}, Shinahr \cite{shinahr1974two} and Szwerinski \cite{szwerinski1982time}.

\subsubsection{Umeo: 2012}

Umeo wrote a survey on solutions to the FSP that use a small number of states\cite{umeo2012recent}. The solutions discussed cover the FSP for one-dimensional arrays, two-dimensional arrays, multi-dimensional arrays and the generalized FSP (GFSP).

The first optimum-time solution to the FSP, developed by Goto in 1962 \cite{Goto1962} had several thousands of states. After that Waksman in 1966 \cite{waksman1966optimum} presented a 16-state optimum-time solution. After that there was a 8-state solution by Balzer in 1967 \cite{Balzer1967}, a 7-state solution by Gerken in 1987 \cite{gerken1987synchronisations} and finally a 6-state solution by Mazoyer in 1987 \cite{Mazoyer1987}. The GFSP has also been extensively studied, and the first optimum-time solution with a small number of states used 17. After that Varshavsky, Marakhovsky and Peschansky in 1970 \cite{VMP1970}, Szwerinski in 1982 \cite{szwerinski1982time}, Settle and Simon in 2002 \cite{settle2002smaller}, Umeo, Hisaoka, Michisaka, Nishioka, and Maeda in 2002 \cite{umeo2003synchronization} presented solution with 10 states and 9 states. There is also a non-optimum-step GFSP solution by Umeo, Maeda, and Hongyo in 2006 \cite{umeo2006design} that works with 6 states.

The paper presents theorems that point that one-dimensional arrays need at least $2n - 2$ steps to synchronize with the general in one of the ends. It also presents a theorem that says that there is no four-state full solution that can synchronize n cells, only four-state partial solution.





\subsection{Maeda et al: 2002}

The contribution of this paper \cite{maeda2002efficient} is a simple but efficient mapping scheme that enables the
embedding any one-dimensional firing squad synchronization algorithm onto two dimensional arrays without introducing additional states. The paper gives a concise and small solution to this problem, where the rules from the 1D array can be easily converted to the 2D array.




In the solution, the authors produce a series of conversion tables and rules to enable the embedding. First they split the m x n cells into groups g. This groups were formed by cells that were on the same line when the m x n board is turned 45 degrees. Because now we have two array lines, we need a state w to be the right and left end state.


As there are more connections between the cells, to convert a cell from one dimension to two dimensions requires more rules. The authors provide a formula for the number of rules that a transformation (a, b, c) $\rightarrow$ d uses depending on the location on the grid.



\section{Open problems}
Although the firing synchronization problem has been studied for decades, there are still several open problems. We discuss open problems for the original problem on the one-dimensional array, for the generalized problem on the one-dimensional array where the initiator can be located in any cell, and for the ring.

\subsection{Original problem}
The most significant open problem in the area is whether there exists a complete 5-state solution to the FSP on a one-dimensional array.

The smallest minimal-time solution so far for the original problem was created by Mazoyer which is a 6-state solution\cite{Mazoyer1987}. A 5-state solution would then optimal for the problem, since Balzer  showed that there is no 4-state minimal-time full solution to the original problem\cite{Balzer1967}, Which later was confirmed by Sanders \cite{Sanders1994}. Some partial 5-state solutions exist. Umeo gives a non-minimal-time partial solution using 5 states\cite{umeo2007smallest}.

Yun\`{e}s has produced solutions based on Wolfram's rules, finding that there exists a 4-state solution to the FSP which synchronizes all lines of length a power of two\cite{yunes20084}. He also proved that there is no 3-state solutions able to synchronize a line of length $n \ge 5$. Various 4-state partial solutions as described by Yun\`{e}s can be found in Table \ref{ovvtable}.

Since all the five-state solutions are partial solutions to the original problem, finding a complete solution would be a significant contribution to the area. It is still unknown whether it is easier to take an approach to the problem finding a minimal-time solution or a non-minimal-time solution. Evidence suggests that finding a non-minimal-time solution may lead to solutions with fewer states, as seen in Table \ref{ovvtable}.

\subsection{The generalized problem}
Before 1998 the best solution for the minimal-time generalized problem was Mazoyer's 10-state solution\cite{Mazoyer1987}. Since then his work has been improved in 1999 by Settle to a 9-state solution\cite{settle1999new}. In 2010 Umeo, Kamikawa and Nishioka \cite{umeo2010generalized} presented a 8-state minimal-time solution which is currently the smallest minimal-time solution to the generalized problem.

Work has also been done on finding non-minimal-time solutions to the generalized problem. In 1999 Settle et al presented a 7-state non-minimal-time full solution to the generalized problem \cite{settle1999new}, which improves the previous minimal-time solution presented by the same authors by two states. In 2006 Umeo, Maeda and Hongyo presented a 6-state non-minimal-time solution\cite{umeo2006design}. This evidence suggests that non-minimal-time solutions may require fewer states than minimal-time solutions, although there is no proof of that claim.

Smaller solutions to the generalized problem are still open problems. Considering that the smallest solution has 6-states and that this number also applies to the original problem, both versions of the problem still do not have a 5-state solution. The 6-state solution to the generalized problem, unlike the 6-state solution to the original problem, is not a minimal-time solution. Because of that, finding a 6-state minimal-time solution to the generalized problem is also an open problem.

\subsection{The ring}

There are some results on state lower bounds for the FSP on the ring.
Berthiaume, Bittner, Perkovic, Settle and Simon showed that there is no 3-state full solution and no 4-state, symmetric, minimal-time full solution to the FSP for the ring\cite{berthiaume2004bounding}. Umeo, Kamikawa and Yun\`{e}s proved that there is no 3-state partial solution to the firing synchronization problem for the ring \cite{umeo2009family}. Thus it is open whether or not there is an unrestricted 4-state minimal-time full solution to the FSP on the ring.





 


\section{Conclusion}
Here we have summarized recent research on the firing synchronization problem, focusing primarily on results published between 1998 and 2015. We discussed results for the original problem on the one-dimensional array, the generalized problem where the initiator can be located in any cell of the array, the problem where the underlying network is a ring, and a paper which discusses how to modify one-dimensional solutions for multidimensional arrays. We also discuss the remaining open problems for the original, generalized, and ring problems.


\bibliographystyle{abbrv}
\bibliography{bibliography.bib}

\end{document}